 \documentclass{paper}

\usepackage[super,compress]{cite}
\usepackage{graphicx}

\usepackage{float}
 \usepackage{amsmath} 
  \usepackage{amssymb}   
\usepackage{multirow} 
\usepackage{cite}
\usepackage{color,xspace}
\usepackage{extarrows}

\numberwithin{equation}{section}

\usepackage{float}
\restylefloat{table}
\allowdisplaybreaks

\newcommand{\gsim}{\lower.7ex\hbox{$\;\stackrel{\textstyle>}{\sim}\;$}}
\newcommand{\lsim}{\lower.7ex\hbox{$\;\stackrel{\textstyle<}{\sim}\;$}}

\newcommand{\nn}{\nonumber}

\title{High frequency gravitational waves from spin-3/2 fields}
\author{\large \bf Karim Benakli \\  \small \textrm {LPTHE, 
Sorbonne Universit\'e et CNRS, } \small \textrm {4 place Jussieu, 75252 Paris Cedex 05, France.}}

\begin{document}

\maketitle

\begin{abstract}
We point out the peculiar form of the gravitational wave signal expected from a gas of particles carry spin 3/2 produced during preheating.  Given the very few ways that gravitinos can manifest themselves in an experimentally observable way, we stress the importance of improving the sensitivity of ultra-high frequency detectors in the future. This review is based on work that appeared in [arXiv:1811.11774 [hep-ph]]\cite{Benakli:2018xdi}.
\end{abstract}

\section{Introduction}

We are used to consistent interacting quantum field theories in flat space-time for particles with spin up to two. Spin one-half (matter: leptons and quarks), one (vector bosons: gauge interactions), two (gravitational interactions, though the gravitons have only been seen in their classical behavior) and zero (the Higgs possibly not a composite state) form the well established basic blocks of our present Standard Models of the Universe. Obviously, in this counting one missing piece is a fundamental particle with spin 3/2. It is tempting then to associate the missing particle with the missing part of the content of the universe: spin 3/2 states represent the main component of dark matter.

A theoretically well motivated spin 3/2 particle is the gravitino and it is in fact one of the first proposed candidates for dark matter\cite{Fayet:1981,Pagels:1981ke,Khlopov:1984pf} (for alternative production mechanism see e.g. \cite{Benakli:2017whb} . This the supersymmetric partner of the graviton and has known consistent description in the framework of supergravity theories.  In gauged supergravities, the charge of the gravitino is related to the value of the cosmological constant making it not suitable for describing our Universe. In flat space-time, the gravitino interacts only through gravitational interactions with the rest of matter. This makes it challenging to detect. There are fortunately a few cases where gravitinos can manifest themselves in a way that can be experimentally observed. The gravitino could be unstable, a very long lived metastable particle if it plays the role of dark matter, and decays. This happens in models with R-parity violation and the main signature is through the emission of neutrinos. Most natural is that the gravitino is stable, in which case we are left with two possible signatures. One is the observation of missing energy at colliders. Relativistic gravitino interactions are dominated by their longitudinal modes. The amplitude of production of the latter is suppressed by the scale of supersymmetry breaking instead of the Planck scale. Finally, one can investigate gravitational signatures of gravitinos. This is the aim of the work \cite{Benakli:2018xdi} reviewed here.

\section{The spin 3/2 fields basics}

The propagating spin 3/2 fields are represented by a spinor-vector representation $\psi_\mu$ and a priori contains too many states. It is constructed as the tensor product:
\begin{align}
(\frac{1}{2},\frac{1}{2}) \otimes (\frac{1}{2},0) =  \frac{1}{2} \oplus (1 \otimes \frac{1}{2}) = \frac{1}{2} \oplus \frac{1}{2} \oplus \frac{3}{2} \ ,
\label{decompose}
\end{align}
where $(a;b)$ with $a,b = 0,\frac{1}{2}$ denote spin-representations of $SU(2)_L \times SU(2)_R$ . The field verifies Dirac equation of motion:
 \begin{eqnarray}
(i{\mathcal{D}} \!\!\!\!/ - m_{3/2}) \psi_\mu &=& 0,  \label{eom}
\label{equations of motion3}
\end{eqnarray}
and two constraints:
 \begin{eqnarray}
M^{\mu \nu}\partial_\nu \psi_\mu &=& 0. \label{constraint2} \\
\gamma_\nu N^{\mu \nu}  \psi_\mu &=& 0,\label{constraint1}
\label{equations of motion3}
\end{eqnarray}
where  the Dirac operator ${\mathcal{D}} \!\!\!\!/$, the matrices $M^{\mu \nu}$ and $N^{\mu \nu}$ are to be determined case by case given the particular space-time and interactions under consideration. The above two constraints allow to project out the non-physical spin 1/2 components. In practice, one can derive the first constraint by noticing that $\psi^0$ appears without time derivative in the Lagrangian; it is a Lagrange multiplier. The other constraint is obtained from the requirement that the first one is conserved in time;  all along only the spin 3/2 component $\psi_\mu$  should propagate.

Working in the Friedmann-Lemaitre-Robertson-Walker (FLRW) space-time, a common trick is to make an appropriate field redefinition that brings the Dirac equation in  to the Minkowski form but with a time-dependent mass. In order to write down the solution of this equation, we introduce the appropriate notation. We define massive spin-1  polarizations $\epsilon^\mu_{\textbf{p}, l}$ satisfying $\epsilon^{\mu}_{\textbf{p}, l} \epsilon^*_{\mu \textbf{p}, l'} = \delta_{l, l'}$. The solution of the Dirac equation can be written in the form:
\begin{equation} 
\mathbf{u}^{(|\lambda|) T}_{\textbf{p}, \frac{s}{2}} (t) = (u_{\textbf{p}, +}^{(|\lambda|)}(t) \chi_s^T(\textbf{p}) ,  \, \,  s \,  u_{\textbf{p}, -}^{(|\lambda|)} (t)  \chi_s^T (\textbf{p})),
 \end{equation} 
$\chi_s (\textbf{p})$ the two-component spinors $\chi_s^\dagger (\textbf{p}) \chi_{s'} (\textbf{p}) = \delta_{s, s'}$ and the time-dependance of the wave function is contained in the scalar wave function $u_{\textbf{p}, \pm}^{(|\lambda|)}(t)$. The solution of (\ref{equations of motion3}) in the momentum space, after Fourier transform on space but not time, are written as  \cite{Auvil:1966eao,Moroi:1995fs,Benakli:2014bpa}
\begin{equation} 
\tilde{\psi}^{\mu}_{\textbf{p}, \lambda}  (t) =  \sum_{s = \pm 1, l = \pm 1, 0} \langle 1, \frac{1}{2}, l, \frac{s}{2} | \frac{3}{2}, \lambda \rangle \epsilon^{\mu}_{\textbf{p}, l} \mathbf{u}^{(|\lambda|)}_{\textbf{p}, \frac{s}{2}} (t),
\label{solRS2}
\end{equation} 
where $  \langle 1, \frac{1}{2}, l, \frac{s}{2} | \frac{3}{2}, \lambda \rangle$ are the Clebsch-Gordan coefficients. 

The equivalence theorem\cite{Fayet:1977vd,Casalbuoni:1988qd,Maroto:1999vd} implies that for energies much higher than the mass of the spin 3/2 particle, the transverse and longitudinal modes decouple from each other, interactions of the latter are those of the would-be goldstino. We can therefore treat separately the two modes. For $\lambda= \pm\frac{3}{2}$, the decomposition (\ref{solRS2}) reads  \cite{Auvil:1966eao,Moroi:1995fs,Benakli:2014bpa}
\begin{equation} 
\tilde{\psi}_{\textbf{p}, \pm \frac{3}{2}}^{\mu } (t) = \epsilon^{\mu }_{\textbf{p}, \pm1}\, \mathbf{u}^{({3}/{2})}_{\textbf{p},\pm \frac{1}{2}} (t).
\end{equation} 
For $\lambda = \pm\frac{1}{2}$, (\ref{solRS2}) reads  \cite{Auvil:1966eao,Moroi:1995fs,Benakli:2014bpa}
\begin{equation} 
\tilde{\psi}_{\textbf{p}, \pm\frac{1}{2}}^{\mu} (t) = \sqrt{\frac{2}{3}}\epsilon_{\textbf{p}, 0}^\mu \,\mathbf{u}^{(\frac{1}{2})}_{\textbf{p} ,\pm\frac{1}{2}} (t) + \sqrt{\frac{1}{3}}\epsilon^\mu_{\textbf{p}, \pm1}\, \mathbf{u}^{(\frac{1}{2})}_{\textbf{p}, \mp\frac{1}{2}} (t).
\end{equation} 
and for $p \gg m_{3/2}$, one can write the expansion $\epsilon_{\textbf{p}, 0}^\mu$,
\begin{equation} \epsilon_{\textbf{p}, 0}^\mu = \frac{1}{m_{3/2}} (p, \sqrt{p^2 + m_{3/2}^2} \hat{\textbf{p}}) = \frac{p^\mu}{m_{3/2}} + \frac{m_{3/2}}{2p} (-1, \hat{\textbf{p}}) + O (\frac{m_{3/2}^2}{p^2}).\end{equation}

The canonical  quantization of the spin 3/2 field starts with the introduction of the set of annihilation and creation operators satisfying the commutation relations: 
\begin{equation} 
\{\hat{a}_{\textbf{p}, \lambda}, \hat{a}_{\textbf{p}', \lambda'}^\dagger\} = (2\pi)^3 \delta_{\lambda, \lambda'} \delta^{(3)} (\textbf{p} - \textbf{p}'). 
\end{equation} 
with
\begin{equation} 
\psi^{(|\lambda|) } (\textbf{x}, t) = \sum_{s =\pm 1} \int \frac{d\textbf{p}}{(2\pi)^3} e^{-i\textbf{p}\cdot\textbf{x}} \{\hat{a}_{\textbf{p}, \lambda} \mathbf{u}^{(|\lambda|)}_{\textbf{p}, \frac{s}{2}} (t) + \hat{a}_{-\textbf{p}, \lambda}^\dagger {\mathbf{v}}^{(|\lambda|)}_{\textbf{p}, \frac{s}{2}} (t) \}.
\label{spinor quantization}
\end{equation} 
The relation between annihilation and creation operators is due to the Majorana nature of the spin 3/2 field considered in this work.

For computing the production of gravitational waves, we will need the energy-momentum tensor  given by:
\begin{eqnarray} 
T_{\alpha\beta} &=& \frac{i}{4} \bar{\psi}_\mu \gamma_{(\alpha}\partial_{\beta)}\psi^\mu - \frac{i}{4}\bar{\psi}_\mu \gamma_{(\alpha} \partial^\mu \psi_{\beta)} + h.c.,\label{stresstensor3/2}
\end{eqnarray}
 and the Hamiltonian
\begin{eqnarray} H(t) 
&=&  \int d\textbf{x}\,  \frac{i}{4} \bar{\psi}^{(\frac{1}{2})} (\textbf{x}, t) \gamma^0 \partial_t \psi^{(\frac{1}{2})}  (\textbf{x}, t) + \frac{i}{4}\bar{\psi}^{(\frac{3}{2})} (\textbf{x}, t) \gamma^0 \partial_t \psi^{(\frac{3}{2})}  (\textbf{x}, t) + h.c., \nn\\
\label{H3}\end{eqnarray}

To make the Hamiltonian diagonal, we will perform a Bogoliubov transformation
\begin{eqnarray} \hat{\tilde{a}}_{\textbf{p}, \lambda} (t) &=& \alpha_{\textbf{p}}^{(|\lambda|) } (t)  \, \hat{a}_{\textbf{p}, \lambda} + \beta_{\textbf{p}}^{(|\lambda|) } (t) \, \hat{a}^\dagger_{-\textbf{p}, \lambda},\nn
\end{eqnarray}
where $p = |\textbf{p}|$ and $\alpha_{\textbf{p}}^{(|\lambda|) } (t)$, $\beta_{\textbf{p}}^{(|\lambda|) } (t)$ are complex numbers satisfying $| \alpha_{\textbf{p}}^{(|\lambda|) } (t) |^2 + |\beta_{\textbf{p}}^{(|\lambda|) } (t) |^2 = 1$. The new time-dependent physical vacuum satisfies
\begin{equation} \hat{\tilde{a}}_{\textbf{p}, \lambda} (t) |0_t\rangle = 0. \label{timedependentvacuum}\end{equation} 

The Hamiltonian reads now 
\begin{equation} H(t) = \int \frac{d\textbf{p}}{(2\pi)^3}  \sqrt{m_{3/2}^2 + p^2} \sum_{\lambda = \pm \frac{1}{2}, \pm \frac{3}{2}}\hat{\tilde{a}}_{\textbf{p}, \lambda}^\dagger (t) \, \hat{\tilde{a}}_{\textbf{p}, \lambda} (t),\label{Hnewbasis}\end{equation} 
and the  the occupation number $n^{(\lambda)}_\textbf{p}(t) = \hat{\tilde{a}}_{\textbf{p}, \lambda}^\dagger (t) \, \hat{\tilde{a}}_{\textbf{p}, \lambda} (t)$  has the expectation value
\begin{eqnarray} 
\langle 0 | n^{(\lambda)}_\textbf{p}(t) | 0 \rangle &=& |\beta_{\textbf{p}}^{(|\lambda|) } (t) |^2\nn\\
  &=& \frac{ \sqrt{m_{3/2}^2 + p^2} - p \, \textrm{Re} (u_{\textbf{p}, +}^{(|\lambda|) *}(t) \, u_{\textbf{p}, -}^{(|\lambda|)}(t)) - m_{3/2}\, (1 - |u_{\textbf{p}, +}^{(|\lambda|)}(t)|^2)}{2 \sqrt{m_{3/2}^2 + p^2}}.\nn\\
  \label{occupationnumber}
  \end{eqnarray}
  
In order to remove the ultraviolet divergence in the momentum integral of the unequal-time correlations, we follow \cite{Enqvist:2012im} and dress the wave functions by a function of the occupation number so that it vanishes at energies above the Fermi surface:
\begin{equation} 
\tilde{u}^{(|\lambda|)}_{\textbf{p},\pm}=\sqrt{2} |\beta_\textbf{p}^{(|\lambda|)}| \, \, u_{\textbf{p},\pm}^{(|\lambda|)}.
\end{equation}

\section{The generic form of the gravitational wave energy spectrum}

The gravitational wave is identified as {\it small perturbations} around the FLRW metric solution of the Einstein equations. It is usual to choose the  transverse-traceless (TT) gauge which leads to a simpler form of the equations. Using the conformal time $\tau$, we have 
\begin{equation} 
ds^2 = a^2(\tau) [-d\tau^2 + (\delta_{ij} + h_{ij}) dx^i dx^j],
\label{FRW}
\end{equation} 
 
It is easier to work in the Fourier space, this allows to avoid non-local operators. We introduce then the TT projection tensor \cite{Weinberg:1972kfs}:
 \begin{equation}  
 \Lambda_{ij, lm} (\hat{\textbf{k}}) \equiv P_{il} (\hat{\textbf{k}}) P_{jm} (\hat{\textbf{k}}) - \frac{1}{2} P_{ij} (\hat{\textbf{k}}) P_{lm} (\hat{\textbf{k}}), \qquad P_{ij} (\hat{\textbf{k}}) = \delta_{ij} - \hat{\textbf{k}}_i \hat{\textbf{k}}_j. 
 \label{defkproj}
 \end{equation} 
 and the ensemble average of the unequal-time correlator of two $\Pi_{ij}^{TT}$
\begin{equation} 
\langle \Pi_{ij}^{TT} (\textbf{k}, t) \Pi^{TT ij} (\textbf{k}', t') \rangle \equiv (2\pi)^3 \Pi^2 (k, t, t') \delta^{(3)} (\textbf{k} - \textbf{k}'),
\label{UTC}
\end{equation} 

Denoting by $\mathcal{P}$ the background pressure, we have:
\begin{equation}
  \Pi_{ij}^{TT} (\textbf{k}, t) = \Lambda_{ij, lm} (\hat{\textbf{k}}) (T^{lm}(\textbf{k}, t) - \mathcal{P} g^{lm} ),
  \label{TTstress}
  \end{equation} 

The TT part of the anisotropic stress tensor $\Pi_{ij}^{TT}$ governs the  production of gravitational waves through:
 \begin{equation} 
 \ddot{h}_{ij} + 2\mathcal{H} \dot{h}_{ij} - \nabla h_{ij} = 16\pi G\Pi_{ij}^{TT},
 \end{equation} 
$\mathcal{H} = \frac{\dot{a}}{a}$ is the comoving Hubble rate. Considering sub-horizon wave lengths, the second term in the equation is sub-leading and will be neglected below.

The spectrum of energy density per logarithmic frequency interval can be expressed as \cite{Dufaux:2007pt}
\begin{equation} 
\frac{d\rho_{ GW}}{d \textrm{log} k} (k, t) = \frac{2 G k^3}{\pi a^4 (t)}\int_{t_I}^t dt' \int_{t_I}^t dt'' a(t') a(t'') {\cos}[k (t' - t'')] \Pi^2 (k, t', t''),
\label{ gravitational wavespectrum}
\end{equation} 
This defines completely the stochastic gravitational background.

The energy domain under consideration is such that $k \gg m_{3/2} \gg \mathcal{H}$. The longitudinal $\lambda, \lambda' = \pm \frac{1}{2}$ and transverse $\lambda, \lambda' = \pm \frac{3}{2}$ modes are typically produced with different rates (see e.g., for gravitinos \cite{Kallosh:1999jj, Giudice:1999yt}) and their production of gravitational should be treated separately. 

We denote
\begin{equation} 
\textbf{p}'= \textbf{p} + \textbf{k}.
\end{equation} 
$\theta$ the angle between $\textbf{k}$ and $\textbf{p}$ and $\theta'$ the angle between  $\textbf{k}$ and $\textbf{p}'$, thus
\begin{equation} 
p' = \sqrt{p^2 + k^2 +2 k p \ {\cos} \theta}, \qquad \theta' = \textrm{arccos} (\frac{p\ {\cos}\theta + k}{ \sqrt{p^2 + k^2 +2 k p\ {\cos} \theta}}).
\end{equation} 

The result of a tedious but straightforward computation of the unequal-time correlation gives\cite{Benakli:2018xdi}
\begin{eqnarray} 
\Pi^2_{|\lambda|} (k, t, t') \simeq \frac{1}{2 \pi^2} \int_{p, p' \gg m_{3/2}} dp\, d\theta \, \, \, \,  K^{(|\lambda|)}(p, k, \theta, m_{3/2}) \, \, \, \tilde{W}^{(|\lambda|)}_{\textbf{k}, \textbf{p}} (t) \tilde{W}^{(|\lambda|)*}_{\textbf{k}, \textbf{p}} (t').\, , \, \, \, \, \, \, \, \, \, \, \, \, 
\label{UTC1/2}
\end{eqnarray}
We have isolated the universal kinematical factors
\begin{eqnarray} 
K^{(\frac{3}{2})} (p, k, \theta, m_{3/2}) &= & p^2 k^2 \{5 \sin^3\theta  \sin^2\theta' +  \sin^2(\theta - \theta')\sin\theta\}
+ 4p^4 \sin^4\theta \sin\theta'.\nn\\
 \label{defK32}
 \end{eqnarray}
and 
\begin{eqnarray} 
K^{(\frac{1}{2})}(p, k, \theta, m_{3/2}) &= &
\frac{1}{36 m_{3/2}^2} p^4p'^2 \sin\theta \{ ({\cos}\theta - {\cos}\theta')^2 + 4 \sin^4 (\frac{\theta-\theta'}{2}) (1 + \sin\theta \sin\theta')\}  \nonumber\\
 &+&   \cdots,
 \label{defK}
 \end{eqnarray}
where the $\cdots$ are sub-leading terms proportional to $m_{3/2}$. 

The model dependent factors:
\begin{equation} 
\tilde{W}^{(|\lambda|)}_{\textbf{k}, \textbf{p}} (t) = 2 |\beta^{(|\lambda|)}_\textbf{p} (t)| |\beta^{(|\lambda|)}_{\textbf{p}'} (t)| \, \{u_{\textbf{p}, +}^{(|\lambda|)}(t) u_{\textbf{p}', +}^{(|\lambda|)}(t) - u_{\textbf{p}, -}^{(|\lambda|)}(t) u_{\textbf{p}', -}^{(|\lambda|)}(t) \},
\label{RegWF}
\end{equation} 
are build from the wave functions solution of the Dirac equation in the particular background and with the sources under consideration in each model.

We can check  that the produced gravitational spectrum is dominated by the helicity 1/2 components as it involves higher power of the momentum $k$.
Plugging (\ref{UTC1/2}) in (\ref{ gravitational wavespectrum}) gives for the leading term:
\begin{equation} 
\frac{d\rho_{ GW}}{d \textrm{log} k} (k, t) \simeq \frac{G k^3}{\pi^3 a^4 (t)} \int dp\, d\theta \,  K^{(\frac{1}{2})}(p, k, \theta, m_{3/2})\ \{| I_{c} (k, p, \theta, t) |^2 + | I_{s} (k, p, \theta, t) |^2\},
\label{ GWspin-3/2 state}
\end{equation} 
with
\begin{equation} 
I_{c} (k, p, \theta, t) = \int_{t_i}^t \frac{dt'}{a(t')} {\cos} (kt') \,\tilde{W}^{(\frac{1}{2})}_{\textbf{k}, \textbf{p}} (t'), \qquad I_{s} (k, p, \theta, t) = \int_{t_i}^t \frac{dt'}{a(t')} \sin (kt') \,\tilde{W}^{(\frac{1}{2})}_{\textbf{k}, \textbf{p}} (t')\, .
\label{spectrum}
\end{equation}

\section{An explicit example}

In our explicit example, we consider gravitinos non-thermally and non-adiabatically produced during preheating. Solving the corresponding Dirac equation leads to explicit forms of the wave functions $u^{(|\lambda|)}_{\textbf{p}, \pm}$. The latter plugged in (\ref{RegWF}) allow then to compute the spectrum of energy density per logarithmic frequency interval from (\ref{spectrum}). The highest amplitude is expected from longitudinal modes; we focus our analysis on their production.

To ensure that we are in the regime $\mathcal{H} \ll m_{3/2}$, we enforce that the supersymmetry breaking by $F$-term in today's vacuum always dominates over the one due to curvature at the end of inflation.  

\begin{figure}[H]
 		\centering
 		\includegraphics[width=0.80\columnwidth]{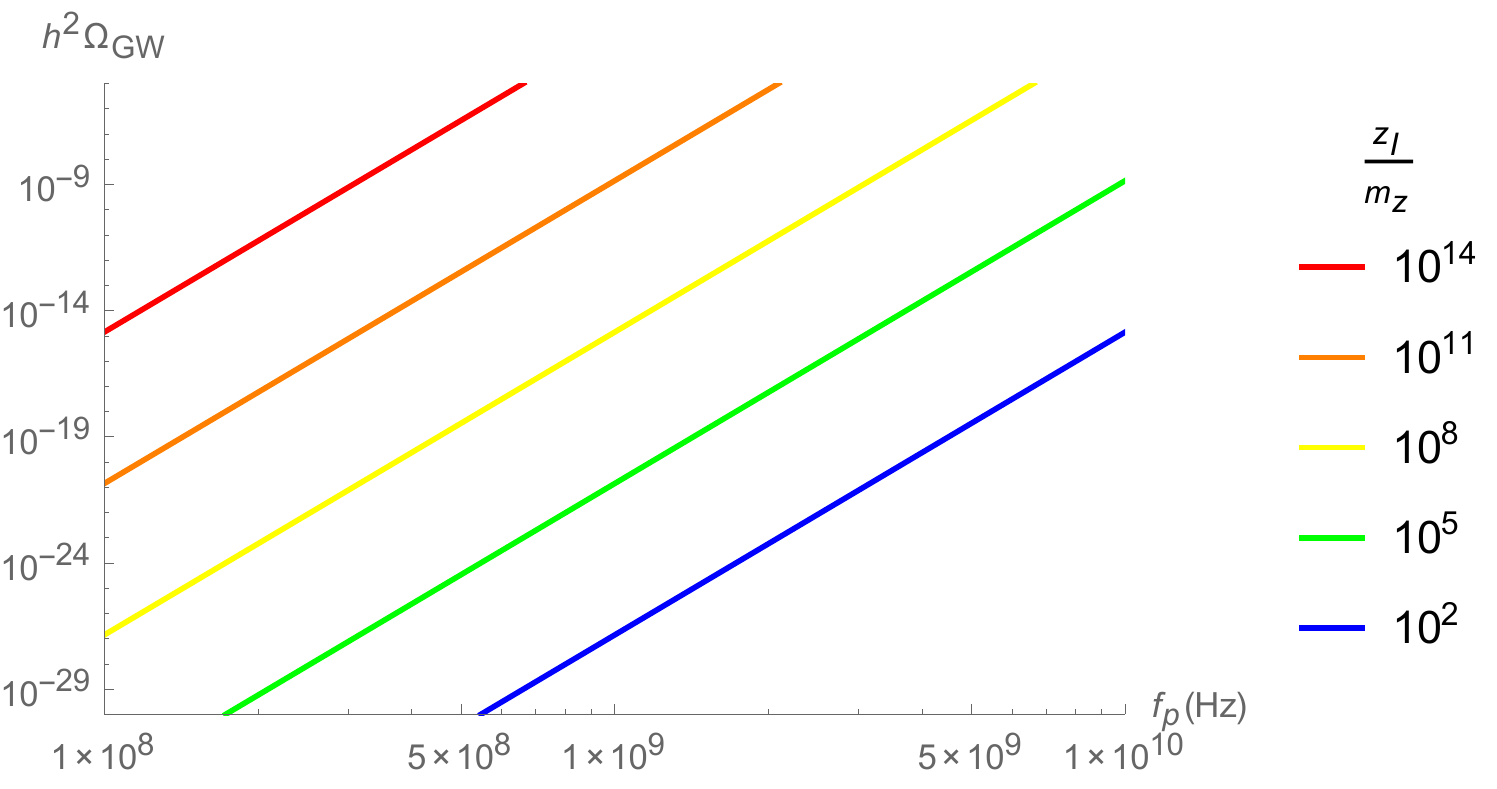}
 		\caption{
 			\footnotesize
 			{The maximal amplitude of  gravitational waves as function of the peak frequency.}
 		}
 		\label{PAplot}
 	\end{figure}

We consider a chiral superfield $z$ with K\"{a}hler potential $\mathcal{K}$ and superpotential $\mathcal{W}$ given by \cite{Ema:2016oxl}.
\begin{eqnarray} 
\mathcal{K} &=& |z|^2 - \frac{|z|^4}{\Lambda^2},\\
\mathcal{W} &=& \mu^2 z + \mathcal{W}_0,
\end{eqnarray}
Near the minimum, we have
\begin{equation} 
m_{3/2} \simeq \frac{\mu^2}{\sqrt{3}M_{Pl}^2} \simeq \frac{\mathcal{W}_0}{M_{Pl}^2}, \qquad m_z \simeq 2\sqrt{3} \frac{m_{3/2} M_{Pl}}{\Lambda}. 
\end{equation} 
at leading order in $M_{Pl}$ and for negligible cosmological constant. We define
\begin{equation}\theta(t) = -\frac{a{m^2_z} {\delta z}}{2\sqrt{3}m_{3/2}M_{Pl}} = -\frac{a{m^2_z} {\delta z}}{2 F},\end{equation}
and $\delta z = z-z_0$ as the displacement of $z$ from its value $z_0$ at the minimum, 
The production of longitudinal modes of the gravitino i.e. goldstinos is described by the Dirac equation  \cite{Kallosh:1999jj, Giudice:1999yt}
\begin{equation}
    [i\gamma^0 \partial_0 - a \, m_{3/2} + (A+iB\gamma^0) \textbf{p} \cdot \mathbf{\gamma}]\begin{pmatrix}u_+ \\ u_- \end{pmatrix} = 0,
\label{eqMotionG}
\end{equation}
where $a$ is the conformal factor in (\ref{FRW}), 
\begin{equation}
 A+iB = \textrm{exp}(2i\int\theta(t)dt), \qquad {\rm and} \qquad f(t)_\pm = \textrm{exp}(\mp i\int\theta(t)dt) u_\pm \, ,
   \label{def equations of motionf}
   \end{equation}
The equation of motion (\ref{eqMotionG}) can be written as
 \begin{equation}
   \ddot{f}_\pm + [k^2+ (a \,m_{3/2} -\frac{a{m^2_z} {\delta z}}{2 F})^2 \mp i \frac{a{m^2_z} {\dot{\delta z}}}{2 F}]f_\pm = 0.
   \label{equations of motion1/2production}
\end{equation}
corresponding to spin-1/2 fermions produced non-adiabatically from the scalar $z$ to which they are coupled through an effective Yukawa coupling 
\begin{equation}
       \tilde{y} = \frac{m_z^2}{2F}.
\label{ytilde}\end{equation} 
The produced fermions fill a Fermi sphere with co-moving radius \cite{Greene:1998nh} $k_F$ and resonance parameter $q$ \cite{Traschen:1990sw} given by
\begin{equation} 
k_F \sim (a/a_I)^{1/4}q^{1/4}m_z, \qquad q \equiv \frac{\tilde{y}^2 z_I^2}{m_z^2}, 
\end{equation} 
where $z_I$ is the initial vacuum expectation value of $z$. The peak of the  spectrum is $k_p \sim k_F$, which corresponds numerically to about
\begin{equation} 
f_p\simeq 6\cdot10^{10} \tilde{y}^{\frac{1}{2}} \textrm{Hz}.
\end{equation} 
The maximum amplitude of the gravitational wave spectrum is estimated to be of order
\begin{eqnarray} 
h^2\Omega_{ GW} (f_p) &\simeq& 2.5 \cdot 10^{-12} (\frac{m_z^2}{z_I M_{Pl}})^2(\frac{a_*}{a_I})^{\frac{1}{2}} q^{\frac{3}{2}} (\frac{k_F}{m_{3/2}})^2 \nn \\
&=& 3 \cdot 10^{-11} \tilde{y}^6 (\frac{z_I}{m_z})^2 (\frac{a_*}{a_I}) \nn \\
&\simeq& 3 \cdot 10^{-10} (\frac{f_p}{6\cdot 10^{10} \textrm{Hz}})^{12} (\frac{z_I}{m_z})^2,
\label{PA}
\end{eqnarray}
where $a_I$ and $a_*$ are the scale factor at initial time and the end of the gravitational wave production. Typical values are displayed in Fig. (\ref{PAplot}) 	
At low frequencies, the spectrum behaves as for a generic gas of spin $1/2$ fields, but near its maximum (\ref{ GWspin-3/2 state}) it scales as $k^5$: this is the very peculiar feature of gravitational wave spectrum produced from spin-3/2 particles.

\section{Conclusion}

The search for fundamental spin 3/2 is hindered by the gravitational nature of their interactions making them some of the most elusive particles in the Universe. We have pointed out that gravitational waves produced from a gas of  spin-3/2 fermions has a peculiar spectrum that can single it out. Though many speculative conditions need to be made so that they are produced in a substantial amount and during preheating, the importance of figuring new ways to observe such new particles can not be underestimated.  On the experimental level, the challenge rests on improving by around five or six order of magnitude the sensitivity of experiments such as in \cite{Sabin:2014bua}.  

\section*{Acknowledgments}
I  would like to thank Y. Chen, P. Cheng and  G. Lafforgue-Marmet for collaboration on this work. This work is supported by the Agence Nationale de Recherche under Grant No. ANR-15-CE31-0002 ``HiggsAutomator''.

\end{document}